\documentclass[aps,pra,twocolumn,showpacs,showkeys,groupedaddress]{revtex4-1}

\usepackage{amsmath,amssymb}

\usepackage[colorlinks=true,bookmarks=false,citecolor=blue,urlcolor=blue]{hyperref} 
\usepackage{color}
\usepackage{graphicx}
\usepackage{mathptmx,courier,textcomp}

\begin{document}

\title{Towards continuous-wave regime teleportation for light matter quantum relay stations}

\author{F. Kaiser$^{1}$}
\email{florian.kaiser@unice.fr}
\author{A. Issautier$^{1}$}
\author{L. A. Ngah$^{1}$}
\author{D. Aktas$^{1}$}
\author{T. Delord$^{2,1}$}
\author{S. Tanzilli$^{1}$}
\email{sebastien.tanzilli@unice.fr}

\affiliation{
$1.$ Universit\'{e} Nice Sophia Antipolis, Laboratoire de Physique de la Mati\`{e}re Condens\'{e}e, CNRS UMR 7336, Parc Valrose, 06108 Nice, France.\\
$2.$ \'Ecole Normale Sup\'{e}rieure de Lyon, 46 All\'{e}e d'Italie, F-69364 Lyon Cedex 07, France.}

\begin{abstract}
We report a teleportation experiment involving narrowband entangled photons at 1560\,nm and qubit photons at 795\,nm emulated by faint laser pulses. A nonlinear difference frequency generation stage converts the 795\,nm photons to 1560\,nm in order to enable interference with one photon out of the pairs, \textit{i.e.}, at the same wavelength. The spectral bandwidth of all involved photons is of about 25\,MHz, which is close to the emission bandwidth of emissive quantum memory devices, notably those based on ensembles of cold atoms and rare earth ions. This opens the route towards the realization of hybrid quantum nodes, \textit{i.e.}, combining quantum memories and entanglement-based quantum relays exploiting either a synchronized (pulsed) or asynchronous (continuous-wave) scenario.
\end{abstract}



\maketitle

\section{Introduction}

Pushing quantum information science one step further will certainly imply augmented compatibility between current quantum technologies, notably those offering pertinent solutions in matter and photonic based quantum systems~\cite{Lvovsky_QM_2009,Tanzilli_Genesis_2012}. This would allow benefiting from the advantages of both worlds and enable true quantum networking applications~\cite{Sangouard_Repeaters_2011}. For example, thanks to the high level of control achieved by experimentalists, atomic and ionic ensembles have been demonstrated to be interesting candidates for storage, manipulation, as well as processing of qubits. On the other hand, photons are ideal qubit carriers for information distribution tasks, as they can propagate over long distance, at high speed, and with essentially no interaction with their environment in both free-space and optical fibers.

One of the main differences between photonic and matter quantum systems lies in their spectral emission and interaction bandwidths, respectively. While typical photonic entanglement sources exhibit spectral bandwidths on the order of some 100\,GHz~\cite{Tanzilli_Genesis_2012}, matter systems are designed for spectral bandwidths ranging from 100\,kHz to 5\,GHz~\cite{Simon_QM_2010}. In addition, matter qubit systems usually operate below 900\,nm, while long distance quantum communication links rather operate best in the telecom C-band of wavelength, \textit{i.e.}, around 1550\,nm where a variety of standard fiber-optic components are available. These severe discrepancies make the compatibility between both worlds very low. For this reason, state-of-the-art long-distance quantum communication devices have so far relied on the generation of photonic qubits without involving matter systems~\cite{Stucki_250km_2009}.

It has been shown that advanced protocols, such as quantum relays based on teleportation and entanglement swapping schemes~\cite{Collins_Swap_2005}, as well as quantum repeater scenarios~\cite{Sangouard_Repeaters_2011}, can allow to further increase the communication distance and the efficiency of quantum networks. However, there are usually two main issues. First, the synchronization of different measurement stations along the communication link becomes a severe limitation at long distances, especially for broadband photons ($\rm \geq 1\,GHz$)~\cite{Riedmatten_Swap_2005,Kaltenbaek_Swap_2009,Aboussan_TPI_2010}. Second, a significant boost in communication speed and distance can only be achieved with quantum memory devices at every relay station, thus requiring interactions between photons and atoms~\cite{Sangouard_Repeaters_2011,Gisin_QC_2007}. Therefore, the future of quantum nodes very likely depends on the ability to realize hybrid quantum systems, coupling standard photonic and matter based devices, having coherently and efficiently matched spectral properties. In this framework the nonlinear optical processes of sum and difference frequency generation are expected to play a more and more important role~\cite{Tanzilli_Interface_2005,Albrecht_Interface_2014}. 

\section{Scope of this paper}

In the following, we describe the realization of a teleportation experiment involving narrowband photons. In our experiment we couple both photons at 1560\,nm for optimal distribution in optical fiber networks and photons at 795\,nm, as obtained from rubidium atomic ensemble based quantum memory devices~\cite{Cho_QM_2010,Specht_QM_2011,Radnaev_QM_2010,Albrecht_Interface_2014}. Our primary goal is to apply the system for a teleportation experiment in which a narrowband photonic qubit, emitted by a matter system (such as a cold atomic cloud) reaches a quantum relay station and its state is teleported onto another photon. In our experiment, we replace the matter based qubit source by faint laser pulses, which allows testing the capabilities of our teleportation scheme with a greater flexibility. The issues associated with the future implementation of true single photon sources based on quantum memories will be discussed in more detail at the end of this paper.

\section{Principle of the teleportation protocol}

The implementation of the teleportation protocol requires two main resources: a single photonic qubit, $|\psi \rangle_{\rm 1}$, and a pair of photonic entangled qubits, $|\psi \rangle_{\rm 23}$. A so-called joint Bell state measurement (BSM) is then performed on the single qubit and one of the entangled qubits. Taking into account the result of the BSM then makes it possible to teleport the original single qubit onto the second of the initially entangled qubits~\cite{Bennett_Tele_1993}.

\begin{figure}[b!]
\centering
\resizebox{1\columnwidth}{!}{\includegraphics{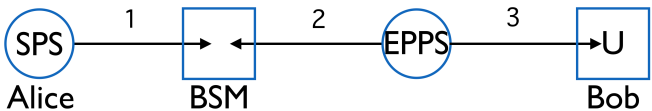}}
\caption{Schematics of the teleportation protocol. Alice and Bob initially share the resource of entanglement, thanks to photons 2 and 3. Alice prepares single qubit states coded on photons 1. Photons 1 and 2 are projected onto an entanglement state at the BSM stage. SPS: single photon source; EPPS: entangled photon pair source.\label{fig_tele}}
\end{figure}

As depicted in \figurename~\ref{fig_tele}, assume Alice prepares, for example, a polarization encoded single qubit of the form
\begin{equation}
	|\psi \rangle_{\rm 1} = \alpha |H \rangle_1 + \beta |V \rangle_1,
\label{Eq_qubit}
\end{equation}
where $|H \rangle$ and $|V \rangle$ denote horizontal and vertical polarization modes, respectively. Here, quantum information is defined by the complex probability amplitudes $\alpha$ and $\beta$, which are normalized as $|\alpha|^2 + |\beta|^2 = 1$. Then, teleportation of $|\psi \rangle_{\rm 1}$ can be achieved by exploiting a pure quantum resource, in this case a maximally entangled state of the form
\begin{equation}
	|\psi \rangle_{\rm 23} = \frac{1}{\sqrt{2}} \left( |H \rangle_2 |H \rangle_3 + |V \rangle_2 |V \rangle_3 \right),
\label{Eq_EntState}
\end{equation}
initially shared by Alice and Bob (see also \figurename~\ref{fig_tele}). For the sake of simplicity, we now assume that all involved photons on which qubits are coded have the same wavelength.
The combined state, $|\psi \rangle_{\rm 123} = |\psi \rangle_{\rm 1} \otimes |\psi \rangle_{\rm 23}$, can be written as
\begin{eqnarray}
	&|\psi \rangle_{\rm 123}& = \, \nonumber \\
	&\,&\frac{1}{2} \bigg( |\Phi^+ \rangle_{12} (\alpha |H \rangle_3 +\beta |V \rangle_3)  \nonumber\\ 
	&\,& + |\Phi^- \rangle_{12} (\alpha |H \rangle_3 - \beta |V \rangle_3) \nonumber \\
	&\,& +|\Psi^+ \rangle_{12} (\beta |H \rangle_3 + \alpha |V \rangle_3)\nonumber\\
    &\,& + |\Psi^- \rangle_{12} (-\beta |H \rangle_3 + \alpha |V \rangle_3)\bigg).
\end{eqnarray}
Here, $|\Phi^{\pm} \rangle_{12} = \frac{1}{\sqrt{2}} \left( |H \rangle_1 |H \rangle_2 \pm |V \rangle_1 |V \rangle_2 \right)$ and $|\Psi^{\pm} \rangle_{12} = \frac{1}{\sqrt{2}} \left( |H \rangle_1 |V \rangle_2 \pm |V \rangle_1 |H \rangle_2 \right)$ are the four maximally entangled Bell states. Consequently, if a Bell state measurement (BSM) is performed on qubits 1 and 2, the latter are projected onto one of those four Bell states, making it possible to retrieve the initial qubit state 1 on Bob's photon 3 after he applies a unitary transformation on his qubit. The unitary transformation to be applied is either $\mathcal{I}_3$, $\sigma_z$, $\sigma_x$, or $\sigma_y$, for a BSM result to be $|\Phi^+ \rangle_{12}$, $|\Phi^- \rangle_{12}$, $|\Psi^+ \rangle_{12}$ or $|\Psi^- \rangle_{12}$, respectively. Here $\mathcal{I}$ stands for the identity operator and $\sigma_i$ for the Pauli operators defined along the three spatial directions ($i\in \left \{x,y,z\right\}$). Note that the teleportation protocol can be extended to that of entanglement swapping as shown in Ref.~\cite{Zukowski_Swapp_1996}.
Also note that the result at the BSM stage not only permits applying the unitary transformation on Bob's qubit 3 but also triggering Bob's detectors conditionally, and therefore augmenting the effective signal-to-noise ratio of the overall quantum channel as expected from quantum relay schemes~\cite{Collins_Swap_2005}.

\section{Experimental realization - qubit source and wavelength conversion}

\begin{figure}[h!]
\centering
\resizebox{1\columnwidth}{!}{\includegraphics{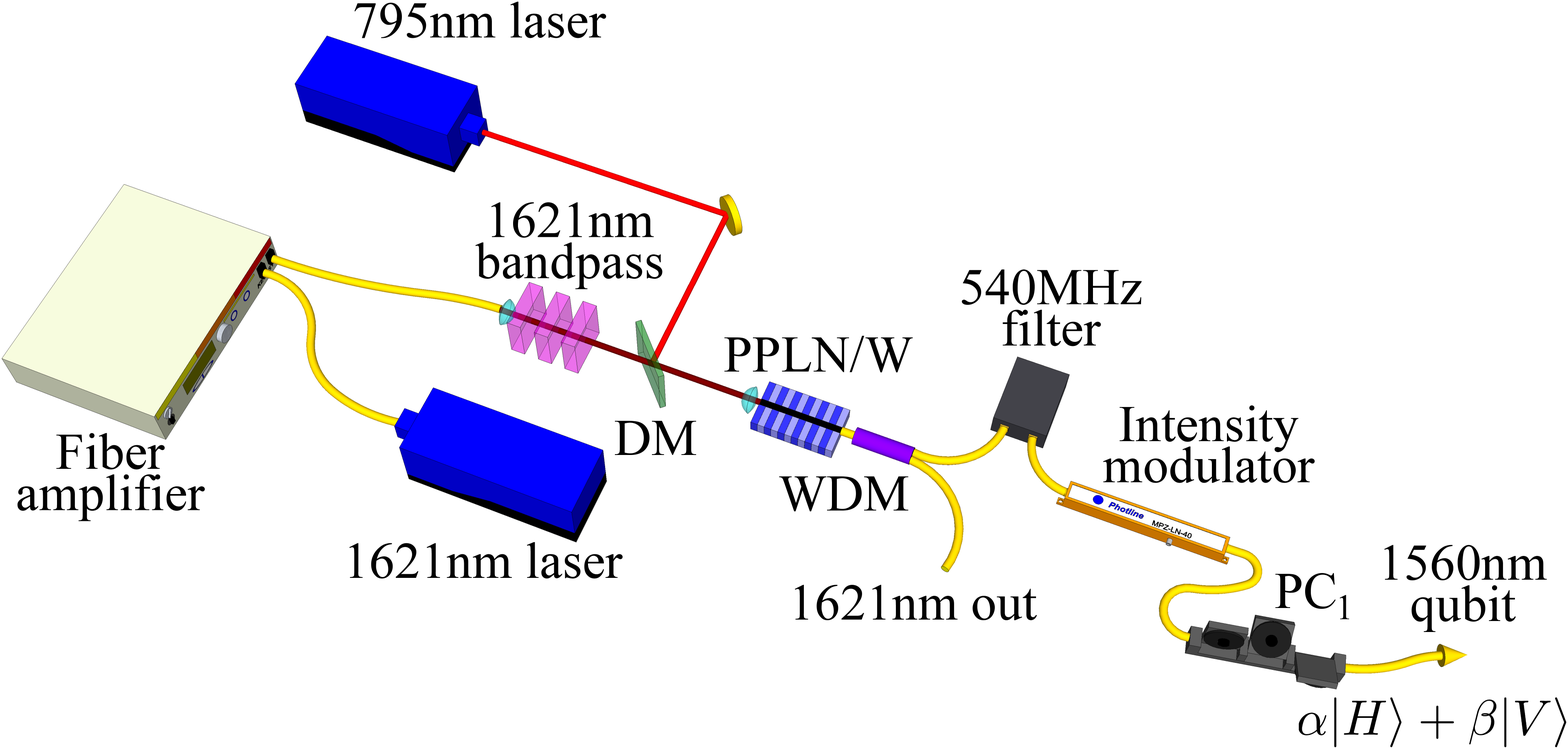}}
\caption{Photonic qubit generator setup. A 795\,nm laser is overlapped with a 1621\,nm at a dichroic mirror (DM) and sent to a PPLN/W. The 795\,nm light is frequency converted to 1560\,nm via DFG. The 1621\,nm pump light induces Raman emission in both the employed optical fibers and the PPLN/W. Several filtering techniques are applied to suppress the emitted Raman photons around 1560\,nm. The photonic qubit is formed using an intensity modulator followed by a polarization controller (PC).\label{fig_QubitSource}}
\end{figure}

\figurename~\ref{fig_QubitSource} shows the realization of the qubit source.

\begin{figure}[b!]
\centering
\resizebox{1\columnwidth}{!}{\includegraphics{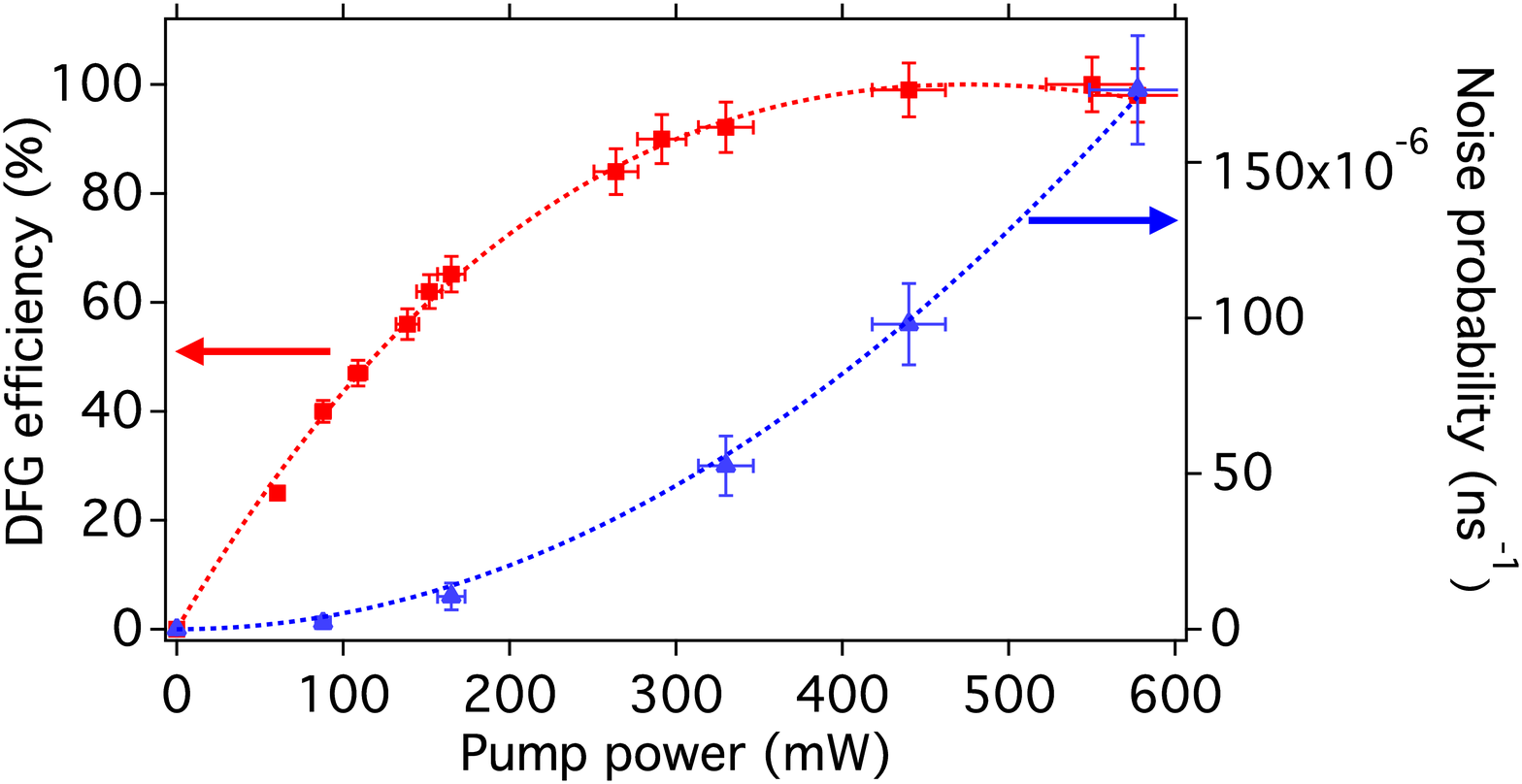}}
\caption{Internal DFG conversion efficiency (red data points and fitting dashed line) as a function of the pump power at 1621\,nm and associated probability of having a noise photon in the wavelength range $\rm 1500 - 1565\,nm$ (blue data points and fitting dashed line). The horizontal error bars assume a 5\%  measurement uncertainty for the employed power-meters. The blue curve shows the probability of having a noise photon, while the vertical ones assume reasonably a poissonnian distribution for the photon detection. 
\label{fig_ConversionAndNoise}}
\end{figure}

We employ a continuous wave 795\,nm laser (Toptica DL pro) to emulate the emission of qubit photons at this particular wavelength, as can be the case, for instance, from an ensemble of cold rubidium atoms~\cite{Bimbard_TomoHSPSRb_2014}. The laser light is then converted to the wavelength of 1560\,nm by means of a coherent quantum interface operated in the difference frequency generation (DFG) regime. Here, by 'coherent' we understand that the interface works for long coherence time photons~\cite{Fernandez_QIntLongPhot_2013}. To do so, we overlap the 795\,nm photons with an intense field at 1621\,nm using a dichroic mirror (DM) and couple both fields into a periodically poled lithium niobate waveguide (PPLN/W) with a length of 3.8\,cm. At a crystal temperature of $\rm 70^{\circ}C$ the phase matching is optimized for the above mentioned DFG process.
In \figurename~\ref{fig_ConversionAndNoise} we show the conversion efficiency from 795\,nm to 1560\,nm as a function of the optical power of the 1621\,nm laser. At around 450\,mW, the optimum conversion efficiency is achieved. Note that the 100\% internal conversion efficiency, as shown in the graph, corresponds to a 27\% overall conversion efficiency which is calculated from the input to the output of the conversion stage, therefore taking into account the losses from the dichroic mirror to the spectral filters.

\begin{figure}[h!]
\centering
\resizebox{1\columnwidth}{!}{\includegraphics{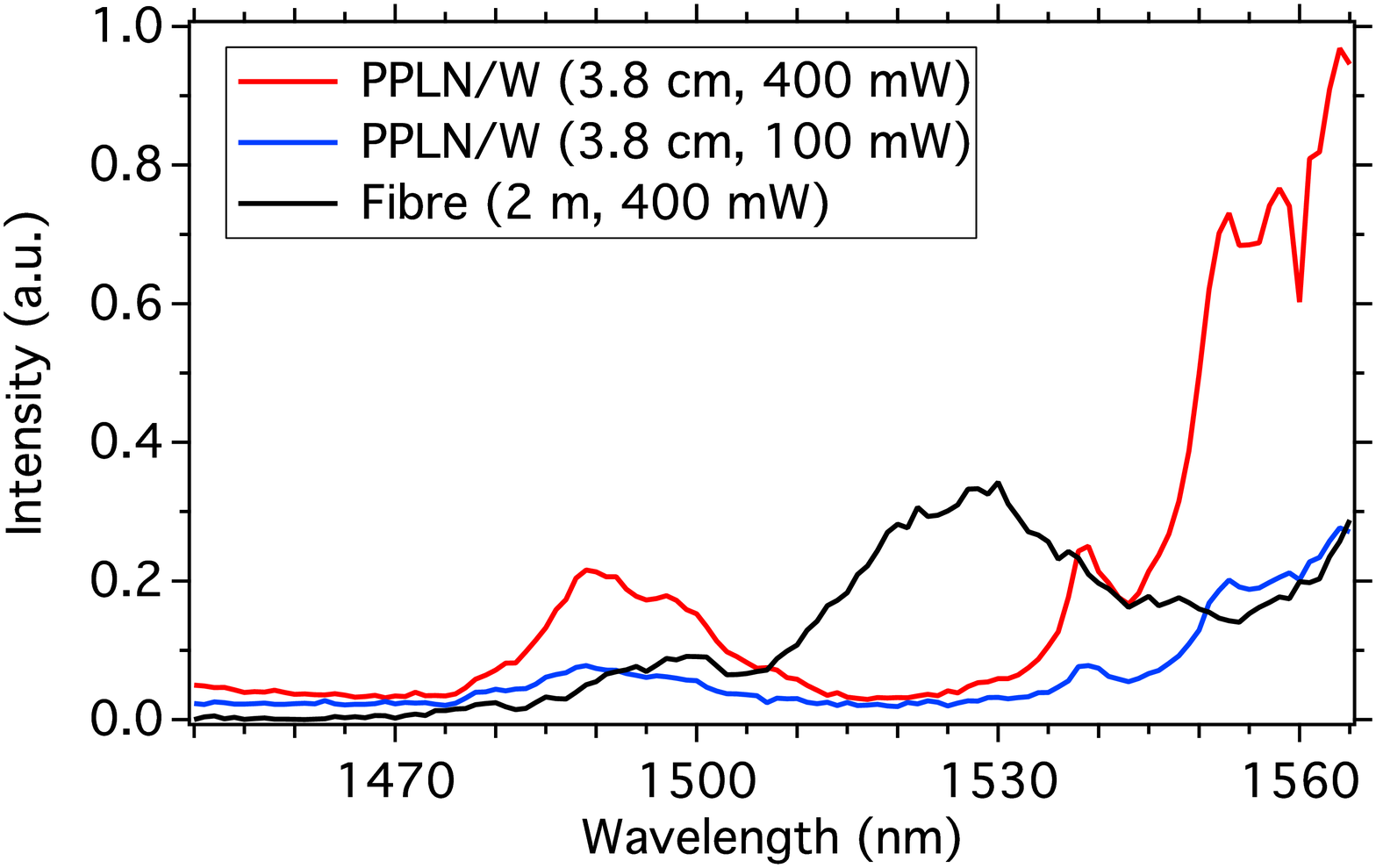}}
\caption{Short wavelength shifted Raman spectrum (anti-Stokes) for both a 2\,m optical fiber and a 3.8\,cm long PPLN/W at different pump powers. Both contributions show undesired Raman noise at 1560\,nm. For the PPLN/W spectrum at 400\,mW (red curve) the dip at 1560\,nm originates from sum frequency generation of the 1621\,nm pump and the 1560\,nm Raman noise ($\rm 1621\,nm\,+\,1560\,nm \rightarrow 795\,nm$). fiber-based Raman scattering can be avoided by coupling light fields into the PPLN/W via free-space.\label{fig_Raman}}
\end{figure}

Similarly as in reference~\cite{Pelc_Raman_2011}, we find additionally that the 1621\,nm light induces a significant noise background due to Raman scattering in the 3.8\,cm long PPLN/W. In addition, we also find strong Raman scattering in optical fibers. To avoid fiber-based Raman scattering noise we chose to couple the 795\,nm and 1621\,nm fields via free-space into the PPLN/W. The measured short wavelength shifted Raman spectrum (anti-Stokes) of both the 3.8\,cm PPLN/W and a 2\,m standard optical fiber (Corning SMF-28e) is shown in \figurename~\ref{fig_Raman}.
We find broadband emission covering more than 100\,nm with a non negligible part of the emission being at 1560\,nm. The blue curve in \figurename~\ref{fig_ConversionAndNoise} shows the noise photon rate in the wavelength range $\rm 1500 - 1565\,nm$ for the PPLN/W only. A quadratic increase in noise is observed with probabilities of about $10^{-4}$ noise photons per nanosecond at near unit conversion efficiency. As this noise contribution is too high for most quantum communication applications, we addressed the problem in three ways. First, fiber Raman noise at the PPLN/W input is removed by using three 1621\,$\pm 5\rm \,nm$ bandpass filters and by coupling light into the waveguide via free space. Second, fiber Raman noise after the PPLN/W is reduced by using a very short collection fiber ($\rm <2\,cm$) followed by a wavelength division multiplexer (WDM) that separates the wavelengths of 1560\,nm and 1621\,nm. To counteract the Raman emission coming from the PPLN/W we use a 1560\,nm narrowband phase shifted fiber Bragg grating bandpass filter after the WDM. It has a spectral bandwidth of about 540\,MHz and essentially reduces the noise rate way below the noise of our detectors ($\rm \sim 10^{-6}\,ns^{-1}$).

In order to generate qubit photons with spectral properties similar to the emission of cold rubidium atomic ensembles~\cite{Bimbard_TomoHSPSRb_2014}, we use the following experimental settings. The 795\,nm laser is operated at a continuous power that corresponds to 0.8 photons per 15\,ns. The DFG module is operated at $\sim 90\%$ conversion efficiency, \textit{i.e.}, we couple $\sim 350\rm \,mW$ at 1621\,nm into the PPLN/W. Additionally accounting for losses in the Raman noise filtering stage leads therefore to an average of 0.2 photons per 15\,ns at 1560\,nm. The continuous wave signal at 1560\,nm is transformed to a 15\,ns pulsed signal using a fast telecom intensity modulator with about 3\,dB of losses. This means that by adjusting the amplitude of the intensity modulator, we can set the probability of having a qubit photon per 15\,ns pulse from 0 to 0.1. In the end, a polarization controller (PC$_{\rm 1}$) is used to code polarization qubit states, $|\psi \rangle_{\rm 1}$, on those photons.

We note that using the intensity modulator after frequency conversion reduces Raman noise counts on the single photon detector by blocking simultaneously the DFG signal, and the Raman noise. Ideally one should employ an intensity modulator at 795\,nm, but we did not have such a modulator at the time of the experiment. We believe that our approach is valid since our Raman noise filtering stage reduces the noise contribution below the dark count level of the detector, even without the intensity modular in place. Consequently, we do not expect significantly different results for employing the intensity modulator before or after the DFG stage.

\section{Experimental realization - teleportation}

\begin{figure}[]
\centering
\resizebox{1\columnwidth}{!}{\includegraphics{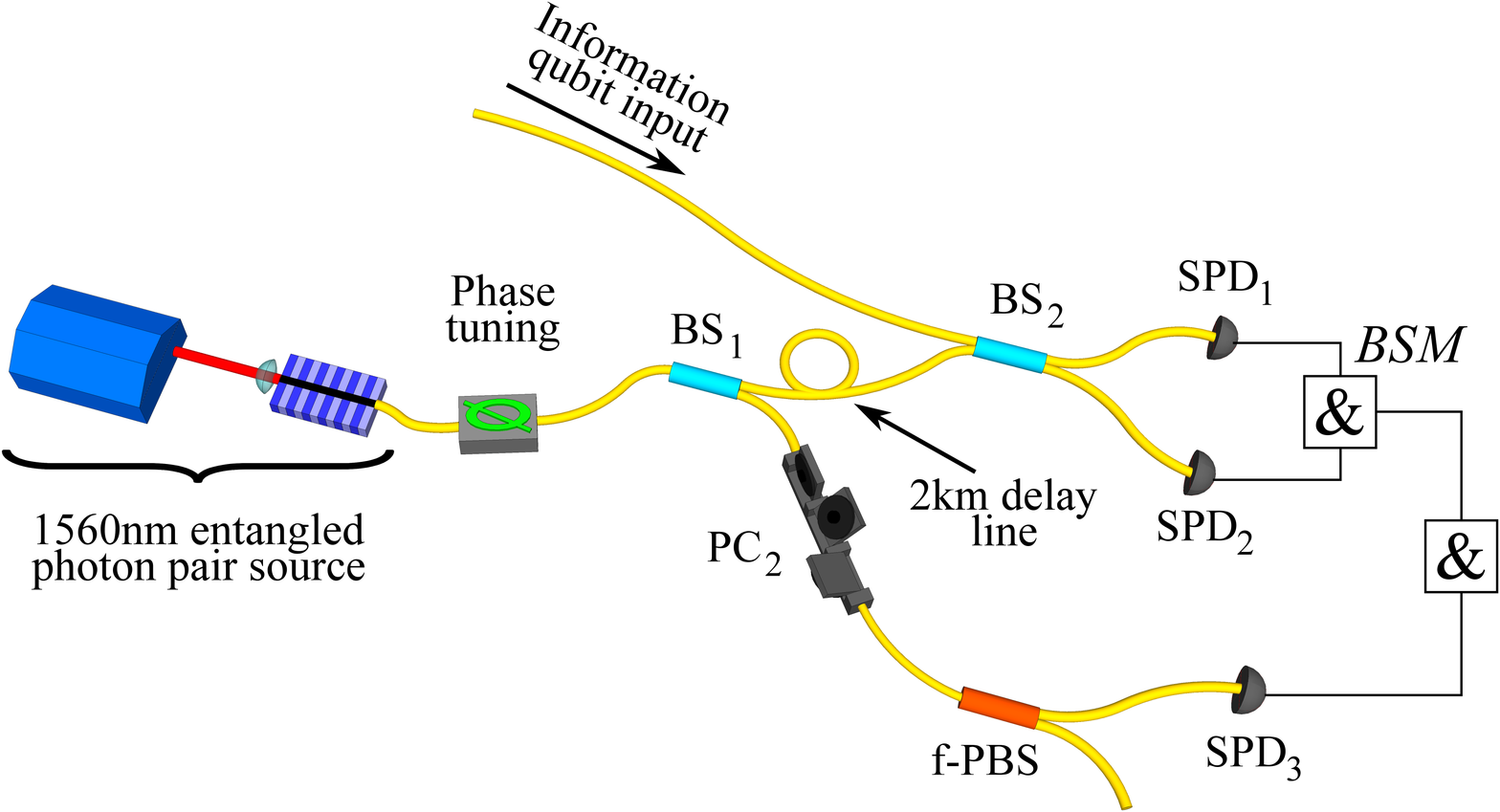}}
\caption{Schematic of the experimental setup. Polarization entangled photons (2 \& 3) are generated in the lower half of the figure and split at BS$_1$. One of the entangled photons is made to interfere with the information qubit at BS$_2$. A BSM is performed using SPD$_1$ and SPD$_2$. The teleported qubit 3 is analyzed using an f-PBS and SPD$_3$ (see text for explanation).\label{fig_Exp}}
\end{figure}

A schematic of the experimental setup for teleportation is shown in \figurename~\ref{fig_Exp}.

As a source of polarization entangled qubit pairs, we use a particularly versatile system based on another high efficiency PPLN/W. The source is based on spontaneous parametric down-conversion (SPDC) and is described in details in references \cite{Kaiser_source_2013,Kaiser_source_2014}. With such a system, high-quality polarization entangled photon pairs at 1560\,nm are generated in the state
\begin{equation}
|\psi \rangle_{23} = \frac{1}{\sqrt{2}} \left( |H \rangle_2 |H \rangle_3 + {\rm e}^{{\rm i}\phi} |V \rangle_2 |V \rangle_3 \right), \label{Eq_SourceState}
\end{equation}
in which $\phi$ is a user controlled phase factor. In addition, the photon spectral bandwidth can be chosen over a large range, namely from 25\,MHz to 4\,THz. For the rest of this paper, the source is operated at 25\,MHz, so as to match with absorption and emission bandwidths of typical atomic ensembles, and also with the qubit source described above. The entanglement source can be operated at rates of $0.01 - 0.03$ photon pairs per coherence time ($\rm \sim 15\,ns$) at the source origin, however, each photon experiences about 10\,dB of loss in the experimental setup.

The BSM between the information qubit and one of the entangled qubits is performed using a fiber optics beam-splitter (BS$_{\rm 2}$) followed by two indium-gallium-arsenide (InGaAs) single photon detectors (SPD). This configuration gives a coincidence between SPD$_1$ and SPD$_2$ only if qubits 1 and 2 are projected onto the fermionic Bell state $|\Psi^- \rangle_{12}$, thus leaving qubit 3 in the state $|\psi \rangle_3 = -\beta |H \rangle_3 + \alpha |V \rangle_3$. In the following we will solely concentrate on this particular event. Whenever the BSM announces the state $|\Psi^- \rangle_{23}$ we use PC$_2$ to perform the necessary unitary operation to transform qubit 3 to the initial state of qubit 1. The state of qubit 3 is analysed using a fiber polarizing beam-splitter (f-PBS) and SPD$_3$.

In an ideal long distance scenario, the qubit source and the entanglement source should be operated independently. Because the photons used in this experiment show coherence times that exceed by far the timing-jitter of the SPDs, this scheme would indeed be achievable~\cite{Halder_ent_2008}. However, due to technical limitations of the intensity modulator driver, we were not capable of generating more than $10^6$ faint laser pulses per second, much lower than the optimal generation rate of $\sim 2.5 \cdot 10^7$ pulses per second. To reduce the measurement times and to increase the probability of having two photons simultaneously at the BSM, we modify the experimental procedure by inverting the measurement order of the teleportation scheme. In other words, as the measurement order in such experiments is irrelevant, at least for fundamental tests~\cite{Kaiser_QDC_2012,MA_FeedTele_2012}, we measure first qubit 3. This announces the presence of qubit 2 which is held in a 2\,km fiber delay line and leaves enough time to herald the generation of qubit 1. The delays are chosen such that qubit 1 and 2 arrive always simultaneously at BS$_2$. This increases significantly the probability of a successful BSM and leads to much shorter measurement times.

\section{Experimental results - characterization}

\begin{figure}[h!]
\centering
\resizebox{1\columnwidth}{!}{\includegraphics{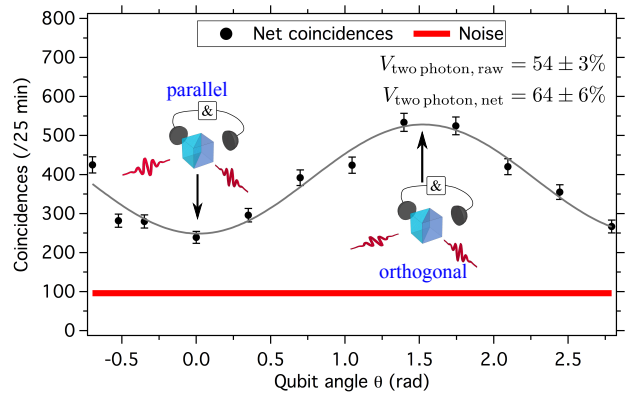}}
\caption{Threefold coincidence rate as a function of the rotation angle $\theta$ applied to the qubit photon. A sinusoidal modulation is obtained and the associated visibilities are in good agreement with the theoretical expectations. This measurement demonstrates that the experimental setup has been properly aligned.\label{fig_PolarHOM}}
\end{figure}

Before performing teleportation, we first characterize the quality of our setup. Since the photonic sources involved in this experiment show non-deterministic photon number statistics, the probabilities of having both multiple qubit photons and multiple entangled qubits at once is non negligible. Such contributions lead to reduced teleportation fidelities, and therefore influence the quality of the experimental results. For characterizing the sources, we perform an initial test using the following conditions. The qubit is prepared as
\begin{equation}
|\psi \rangle_{\rm 1} = \sin \theta |H\rangle_1 + \cos \theta |V \rangle_1,
\end{equation}
in which $\theta$ is a polarization rotation angle, set by PC$_1$ (see \figurename~\ref{fig_QubitSource}). Qubit 3 is analysed as $|H \rangle_3$ and the threefold coincidence rate between SPD$_1$, SPD$_2$ and SPD$_3$ is measured as a function of $\theta$. The optical powers are set such that we have a probability of 0.02 generated photon pairs per 15\,ns (at the PPLN/W output), and 0.03 photons per laser pulse (in front of BS$_2$).
The result is shown in \figurename~\ref{fig_PolarHOM}. As expected, a sinusoidal modulation of the threefold coincidence rate is obtained.
To infer the quality of this measurement, we define visibility parameter similar to that used in two-photon interference type experiments related to teleportation or entanglement swapping protocols~\cite{Martin_NJP_12,Halder_NatPhys_2007}
\begin{equation}
\label{Eq_V_2phot}
	V_{\rm\,two\,photon} = \frac{C_{\rm max} - C_{\rm min}}{C_{\rm max}},
\end{equation}
in which $C_{\rm max}$ and $C_{\rm min}$ denote the maximum and minimum coincidence rates, respectively. We obtain a raw visibility of $54\pm3\%$, and correction for false events originating from detector dark counts leads to a net visibility of about 64\% (see appendix A for details on dark count subtraction).

\begin{figure}[b!]
(a)\\
\resizebox{1\columnwidth}{!}{\includegraphics{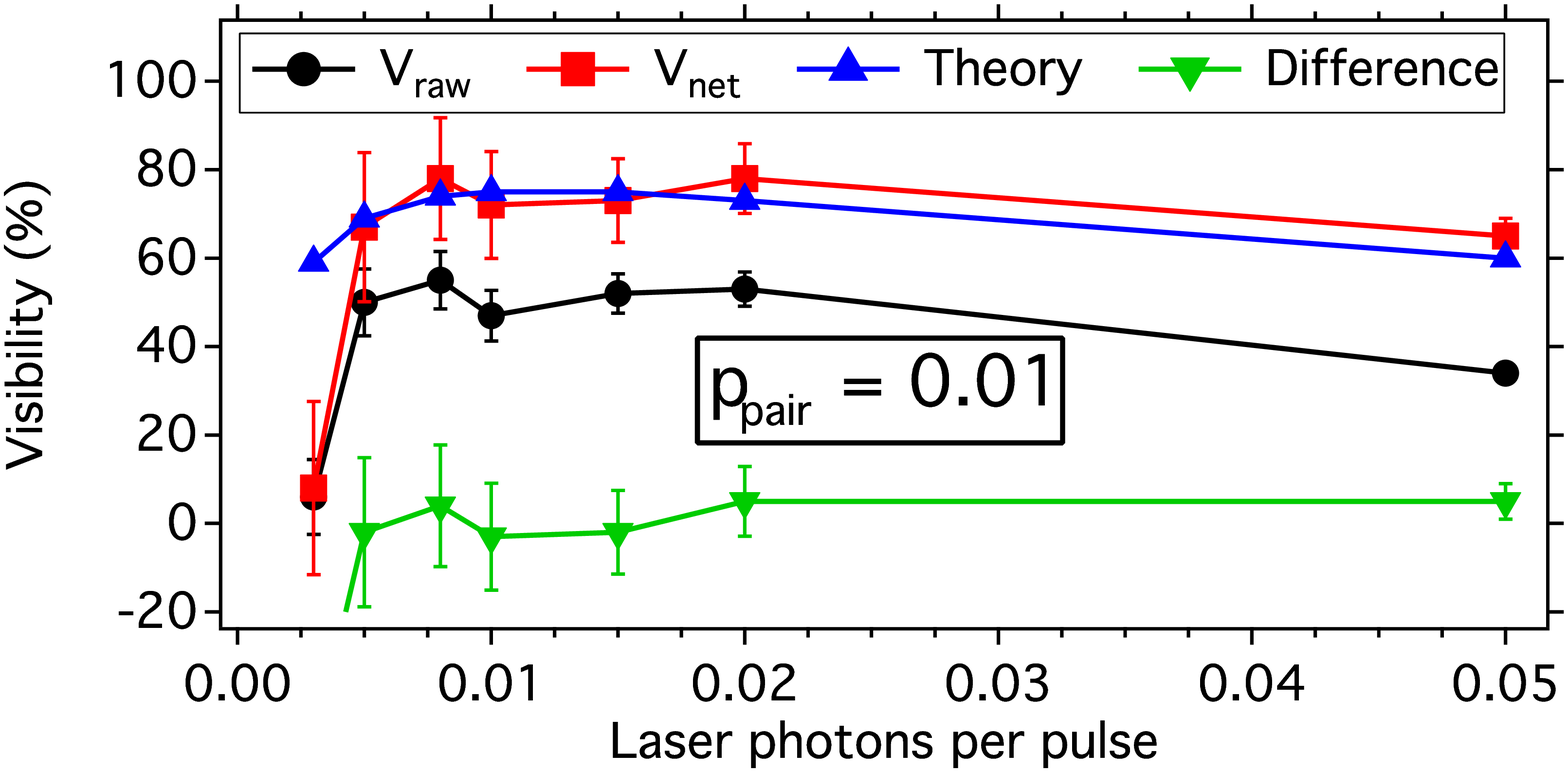}}\\
(b)\\
\resizebox{1\columnwidth}{!}{\includegraphics{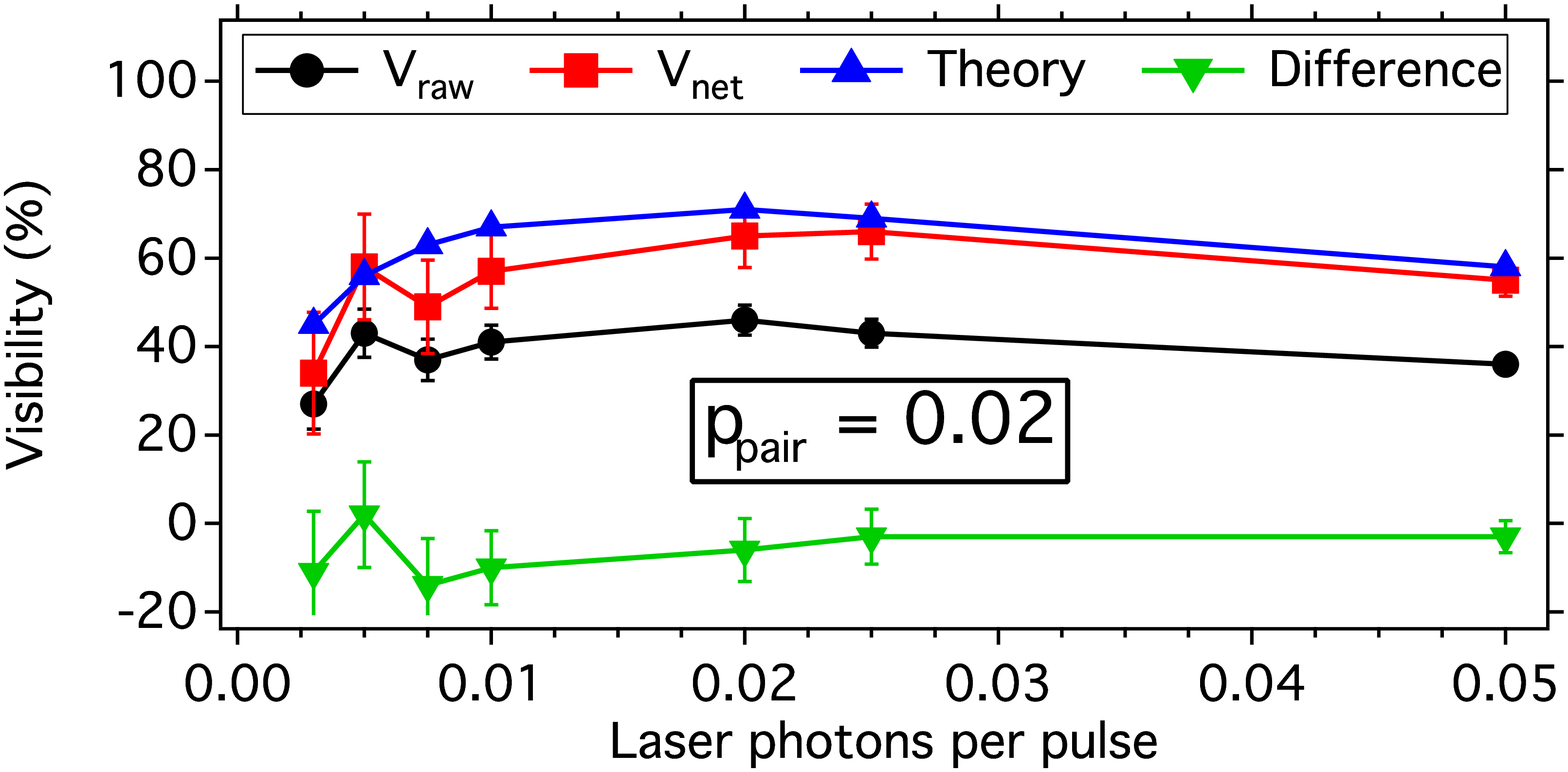}}\\
(c)\\
\resizebox{1\columnwidth}{!}{\includegraphics{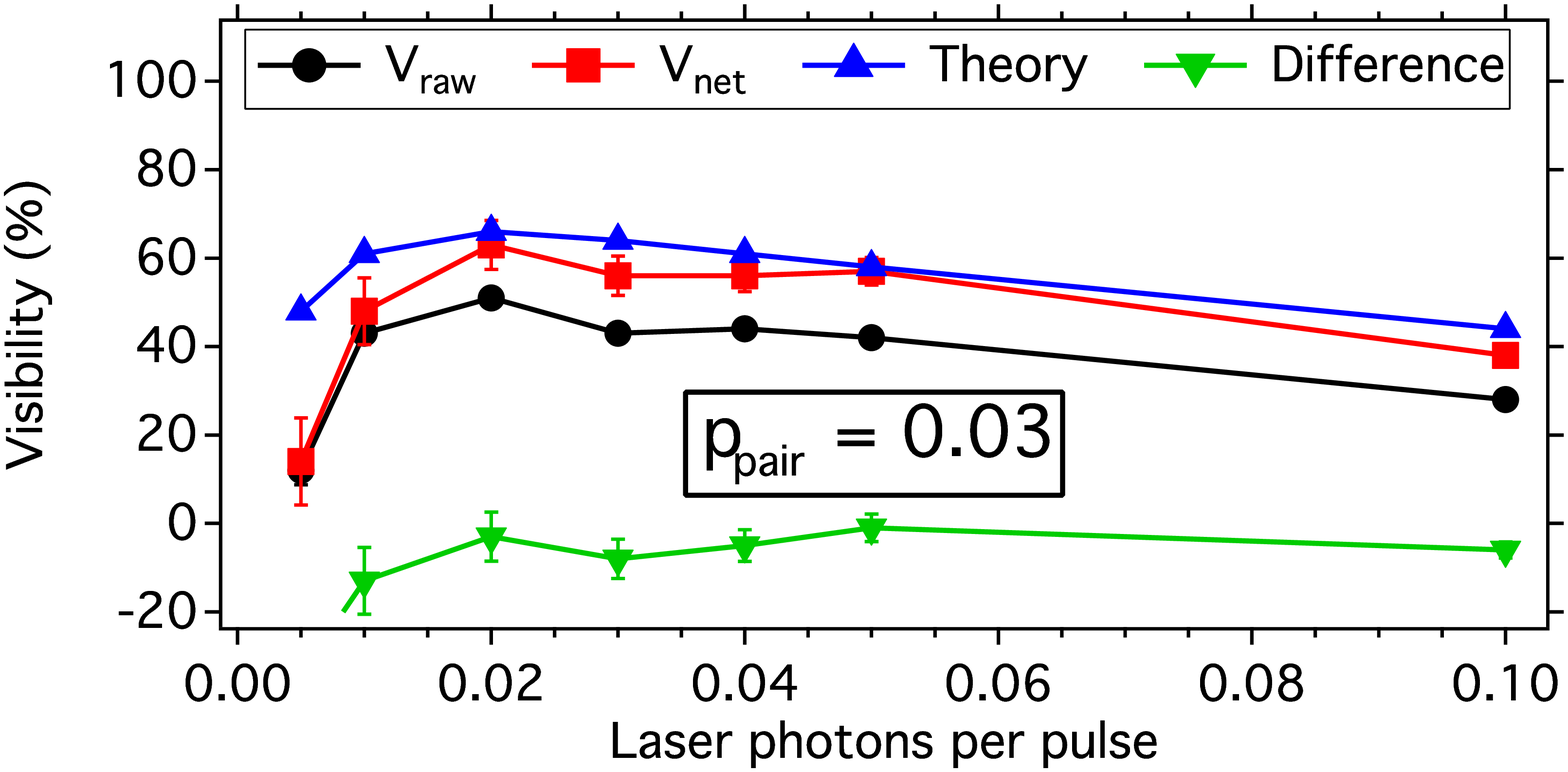}}
\caption{Obtained visibilities for probabilities of having a single photon in a qubit laser pulse. From (a) to (c), the probability of generating an entangled photon pair per coherence time, $p_{\rm pair}$, is increased from 0.01 to 0.03. Raw (net) visibilities of up to 55\% (75\%) are obtained. In all graphs, the net visibilities agree reasonably well with the theoretical expectations (within about 10\%). The green points indicate the difference between raw and theoretical visibilities. All lines between data points are guides to the eye. The comparatively strong differences for low numbers of laser photons per pulse can be explained by the following reasons. Due to noise from the DFG stage it was hard to precisely characterise the real number of qubit photons per laser pulse. Additionally, there is a very strong dependence of the visibility as a function of the photon number per pulse in this region, such that small intensity fluctuations cause a strong error in visibility.\label{fig_PumpPower}}
\end{figure}

\begin{figure}[h!]
\centering
\resizebox{1\columnwidth}{!}{\includegraphics{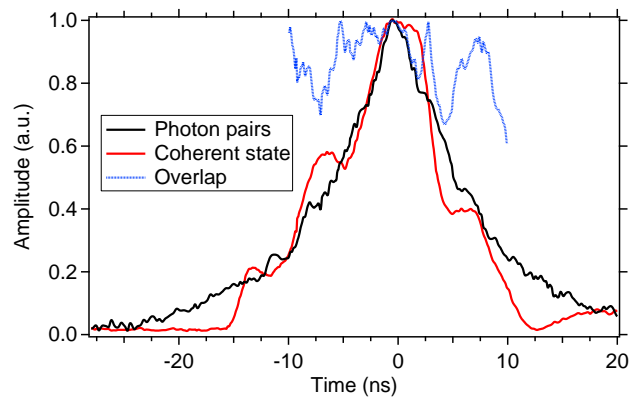}}
\caption{Temporal shape of photons obtained from the SPDC photon-pair source (black) and attenuated laser pulses (red). The overlap between both signals is shown as blue dashed line. The overlap integral in the region of interest ($\pm4\rm \,ns$) is about 91\%.\label{fig_Overlap}}
\end{figure}

To characterize our photonic sources, and setup in a more general way, we repeat this measurement for various mean numbers of photons. We tune the entangled photon pair generation from 0.01 to 0.03 pairs per 15\,ns and the laser power from 0.0025 to 0.1 photons per pulse.
\figurename~\ref{fig_PumpPower} represents the obtained and related visibilities.
We can attain a maximum raw visibility of about 55\%, and correction for dark count events leads to a net visibility of about 75\%.
Aside from non perfect photon number statistics, non perfect visibilities are explained by two main limitations. First, as shown in \figurename~\ref{fig_Overlap}, the temporal shapes of the qubit photon and entangled photons show an overlap of only 91\% in the integration region ($\pm4\rm \,ns$), due to the capabilities of the intensity modulator driver. In addition, the entangled photon pairs experience losses of about 10\,dB and are split in a non deterministic fashion at BS$_1$, which introduces another reduction in the maximum visibility. Taking into account these experimental imperfections, we obtain a reasonably good agreement between theory and experiment (see appendix B for more details).

Note that these initial visibility tests have been performed in the so-called phase-insensitive $\{H;\,V\}$ basis, which does not represent a proof of teleportation, but rather a quality test of the optical setup. In other words, a purely classical state would lead to the same results. In order to prove the quantum nature of our observations we need to perform an experiment in the phase-sensitive basis in which control on the coherence of the employed states matters.

\section{Experimental results in the teleportation regime}

To demonstrate teleportation in the phase sensitive basis, there are three options. One could either perform a quantum state tomography measurement via generating several different input qubit photons and comparing them with the measured qubit 3 state. Such a procedure is a tedious work requiring long measurement times for polarization qubits. Alternatively, one could repeat the measurement of \figurename~\ref{fig_PolarHOM} and analyse qubit 3 in the diagonal basis $\{ D \}$, where $|D \rangle_3 = \frac{1}{\sqrt{2}} (|H \rangle_3 + |V \rangle_3)$. However, in our particular fully fiber setup, the phase relation $\phi_3$ between the contributions $|H \rangle_3$ and $|V \rangle_3$ could not be determined. Consequently, we choose to perform energy-time like measurements to demonstrate teleportation in the phase sensitive basis. This allows simplifying the experimental implementation, as the exact value of $\phi_3$ becomes unimportant as long as it is stable during the measurement times. Note that the related theoretical background is given in appendix C. In short, qubit 1 (see equation~\ref{Eq_qubit}) has to be prepared in the diagonal state of polarization, \textit{i.e.} $\alpha = \beta = \frac{1}{\sqrt{2}}$ (or $\theta = 45^{\circ}$). Moreover, the phase $\phi$ in the entangled state (see equation~\ref{Eq_SourceState}) is scanned and qubit 3 is analyzed in the diagonal basis. As a consequence, we expect a sinusoidal dependence of the coincidence rate as a function of $\phi$. The fringe visibility (V), now defined as
\begin{equation}
\label{Eq_V_ent}
V_{\rm ent} = \frac{C_{\rm max} - C_{\rm min}}{C_{\rm max} + C_{\rm min}},
\end{equation}
as usually done for entanglement based measurements, quantifies then the teleportation fidelity (F) as $F = \dfrac{1+V_{\rm ent}}{2}$~\cite{Marcikic_Tele_03,Riedmatten_Swap_2005}. Compared to the full quantum state tomography this measurement requires much less measurements and can therefore be performed in a much shorter time.

\begin{figure}[h!]
\centering
\resizebox{1\columnwidth}{!}{\includegraphics{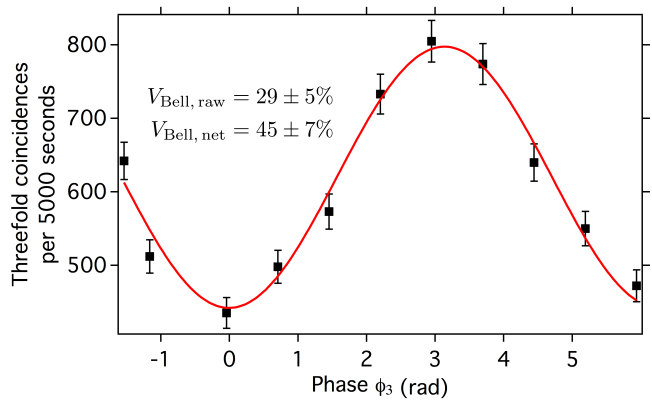}}
\caption{Experimental results for the energy-time like teleportation experiment. A sinusoidal coincidence rate modulation is obtained as a function of the phase $\phi$ in the entangled state. The red line is a sinusoidal fit to the data. \label{fig_results}}
\end{figure}

The experimental results for a photon pair generation rate of 0.02 pairs per 15\,ns, and 0.02 photons per laser pulse are shown in \figurename~\ref{fig_results}.
We obtain a raw fringe visibility of $29\pm5\%$ and a net visibility of about 45\% which is in good agreement with the theoretical expectation.

\section{Conclusion and outlook}

We have demonstrated a teleportation experiment in which narrowband photonic qubits at 1560\,nm are coupled at a Bell state measurement stage. Single photons emulated from a faint laser operating at 795\,nm, were converted to 1560\,nm via the non-linear process of difference frequency generation in a periodically poled lithium niobate waveguide. They were converted to pulses using an intensity modulator and then measured in a joint manner with photons out of entangled pairs. This teleportation experiment paves the way towards continuous wave regime teleportation experiments with single photons emitted by atomic quantum memories. In this case the observed interference visibilities could also be strongly improved by better photon number statistics. One of the main limitations of the current experiment is the poissonian photon statistics for the single qubit photons.

Regarding the non-linear conversion stage, we have outlined strategies for reducing the Raman scattering induced noise at 1560\,nm, so as to obtain essentially noise free photon conversion. Moreover, the resource of photonic entanglement was obtained from a highly versatile source that permits to generate polarization entangled photons over a bandwidth of 25\,MHz, but also to tune the phase between the two contributions to the entangled state.

Concerning the performance of the reported teleportation scheme, the obtained raw and net visibilities lead to corresponding fidelities that are, strictly speaking, below the cloning threshold (5/6) for inferring a true quantum teleportation~\cite{Scarani_Cloning_2005}. However, we note that this is mainly due to technical reasons which can be overcome with current technologies. The limited speed of the intensity modulator and associated driver ($\rm\sim 150\,MHz$) reduces the temporal and spectral overlap between the single and entangled photons to about 91\%. A faster modulator and an arbitrary waveform driver would overcome this problem. In addition, visibilities could be increased by deterministically separating the entangled photon pairs, as demonstrated in references~\cite{Halder_ent_2008,Kaiser_typeII_2012}. Improved visibilities would also be achieved by reducing the propagation losses, \textit{e.g.} by splicing all fiber connections. Finally, and this is the main contribution, employing a heralded single photon source, as those based on an emissive quantum memory, would allow to obtain near unit net visibilities~\cite{Albrecht_Interface_2014,Bimbard_TomoHSPSRb_2014}. Note that for a real application scenario, the raw visibility stands as the figure of merit of interest, implying the use of much less noisy single photon detectors, \textit{e.g.} those based on superconducting nanowires~\cite{Hadfield_Detectors_2009}. By applying all the above mentioned modifications, demonstrations of long distance quantum relay scenarios employing both light and matter systems could be achieved in the near future.

\appendix

\section{Measurement of dark count contributions}

To measure the detector dark counts induced coincidences in our experiment we use the following strategy. We operate both the SPDC photon pair source and the coherent light source at nominal settings. Then we measure the dark count rates DC$_1$, DC$_2$, DC$_3$, DC$_{12}$, DC$_{13}$, and DC$_{23}$. Here the subscripts denote the respective detectors that are blocked. The total dark count rate DC is then
\begin{equation}
{\rm DC} = {\rm DC_1} + {\rm DC_2} + {\rm DC_3} - {\rm DC_{12}} - {\rm DC_{13}} - {\rm DC_{23}}.
\end{equation}

\section{Calculation of the maximal expected visibility}

Both the photon pair source and the coherent source show non ideal photon number statistics, \textit{i.e.} they emit occasionally more than one photon pair, or more than one qubit photon at a time. This reduces the teleportation fidelity because the interference visibility at BS$_2$ is reduced for multiphoton inputs. It is therefore necessary to operate the sources at very low photon fluxes to minimize multiphoton effects, typically on the order of $0.01 - 0.1$ generated photons (or photon pairs) per coherence time.

In addition, losses in the experimental setup and non unit detection efficiencies reduce the probability of detecting, at the same time, three photons on SPD$_1$, SPD$_2$ and SPD$_3$.

In the following we will outline the basic ideas of how to calculate the reduced visibilities when accounting for these imperfections.\\
1.) The state emitted by the entangled photon pair source can be approximated by
\begin{equation}
|\psi \rangle_{\rm 23} \propto \sqrt{p_0} \, |0 \rangle + \sqrt{p_1} \, |2 \rangle + \sqrt{p_2} \, |4 \rangle + \mathcal{O},
\end{equation}
in which $p_0$, $p_1$, and $p_2$ denote the probability of generating zero, one or two photon pairs per coherence time. The numbers in the ket vectors denote the photon numbers.
Usually, the probabilities $p_0$ and $p_2$ are expressed as function of $p_1$:
\begin{eqnarray}
p_0 &=& 1-p_1-p_1^2 \\
p_2 &=& p_1^2.
\end{eqnarray}
In a similar fashion, the state emitted by the coherent source is
\begin{equation}
|\psi \rangle_{\rm 1} \propto \sqrt{l_0} \, |0 \rangle + \sqrt{l_1} \, |1 \rangle + \sqrt{l_2} \, |2 \rangle + \mathcal{O},
\end{equation}
in which $l_0$, $l_1$, and $l_2$ denote the probabilities of having 0, 1, or 2 qubit photons per coherence time, \textit{i.e.} within a laser pulse. For our laser source we have also
\begin{eqnarray}
l_0 &=& 1-l_1-\frac{l_1^2}{2} \\
l_2 &=& \frac{l_1^2}{2}.
\end{eqnarray}\\
\,\\
2.) The photons from the SPDC source propagate towards BS$_1$ and we can compute the probabilities for the eight possible outcomes $|4\rangle_2 |0 \rangle_3$, $|3\rangle_2 |1 \rangle_3$, $|2\rangle_2 |2 \rangle_3$, $|1\rangle_2 |3 \rangle_3$, $|0\rangle_2 |4 \rangle_3$, $|2\rangle_2 |0 \rangle_3$, $|1\rangle_2 |1 \rangle_3$, and $|0\rangle_2 |2 \rangle_3$.
The subscripts 2 and 3 denote the two possible paths at the output of BS$_1$.\\
\,\\
3.) The probability that the photons in path 3 give a detection event is given by
\begin{equation}
P(N,\,t_3,\,\eta) = 1-(1-t_3\,\eta)^N,
\end{equation}
where $N$ is the number of photons in path 3, $t_3$ is the transmission of path 3, and $\eta$ the detection efficiency of the non photon number resolving detector. With these tools we can then calculate the probabilities of heralding 0, 1, 2, or 3 photons in path 2. For calculating the probabilities $h_0$, $h_1$, $h_2$, and $h_3$ of having 0, 1, 2, or 3 photons in front of BS$_2$ we need also to consider the transmission of path 2, $t_2$.\\
\,\\
4.) Since we are dealing with interference on BS$_2$ we need to transform the probabilities $h_0$, $h_1$, $h_2$, and $h_3$ to probability amplitudes by taking the square root. At the other input port of BS$_2$ we have the photons from the coherent source in the state $|\psi \rangle_{\rm 1}$. The probability amplitudes for all possible outcomes at BS$_2$ (ranging from $|5\rangle_1 |0 \rangle_2$ to $|0\rangle_1 |5 \rangle_2$) are then computed for distinguishable and indistinguishable photons. Squaring the absolute value of these probability amplitudes and accounting for the respective detection efficiencies (via multiplication with $P(N,\,1,\,\eta)$) allows us to compute the rate of threefold detection events, $C_{\rm dis} \equiv C_{\rm max}$ and $R_{\rm indis} \equiv C_{\rm min}$, heralded by a click on detector SPD$_3$.\\
\,\\
5.) The maximum attainable visibilities $V_{\rm two\,photon}$ and $V_{\rm ent}$ can then be calculated as a function of $p_1$, $l_1$, $t_1$, $t_2$, and $\eta$.

\begin{figure}[h!]
\centering
\resizebox{1\columnwidth}{!}{\includegraphics{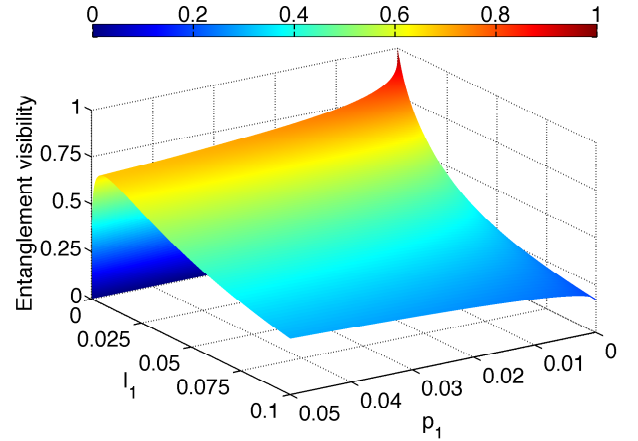}}
\caption{Maximum achievable entanglement visibility $V_{\rm ent}$ for imperfect photonic sources. The probability of generating, via SPDC, one photon pair per coherence time is $p_1$. The probability having a photon from the coherent source per coherence time is $l_1$. The colorbar on top indicates the visibility.\label{fig_TheoVisi}}
\end{figure}

\begin{figure}[h!]
\centering
\resizebox{1\columnwidth}{!}{\includegraphics{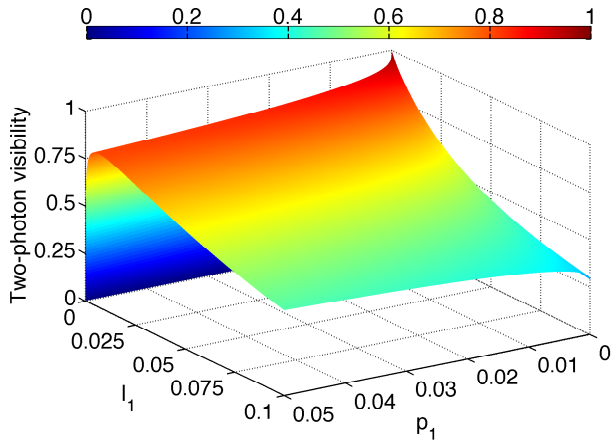}}
\caption{Maximum achievable two-photon interference visibility $V_{\rm two\,photon}$ for imperfect photonic sources. The probability of generating, via SPDC, one photon pair per coherence time is $p_1$. The probability having a photon from the coherent source per coherence time is $l_1$. The colorbar on top indicates the visibility.\label{fig_TheoVisi2}}
\end{figure}

An example, the maximum achievable visibilities (two-photon interference, and entanglement) as a function of $p_1$ and $l_1$ are shown in \figurename~\ref{fig_TheoVisi} and \figurename~\ref{fig_TheoVisi2}. For these plots we assume $t_1 = t_2 = 0.1$ and $\eta = 0.2$.

\section{Theoretical background for energy-time like teleportation}

To demonstrate that our experiment indeed performs teleportation, we need to show that the experimental settings at the single qubit generator and the qubit analysis stage influence the threefold coincidence rate in a nonlocal fashion.
We derive in the following the theoretical framework for our experiment and describe afterwards our particular configuration.

The single qubit is prepared in the state:
\begin{equation}
 |\psi \rangle_1 = \frac{1}{\sqrt{2}} \left( { \rm e}^{{\rm i} \frac{\phi_1}{2}}  |H \rangle_1 + { \rm e}^{-{\rm i} \frac{\phi_1}{2}}  |V \rangle_1  \right),
\end{equation}
in which $\phi_1$ is a relative phase factor between the two contributions to the state. 
Moreover, the entangled qubit pair is prepared in the state
\begin{equation}
|\psi \rangle_{23} = \frac{1}{\sqrt{2}} \left( {\rm e}^{{\rm i}\frac{\phi}{2}} |H \rangle_2 |H \rangle_3 + {\rm e}^{-{\rm i}\frac{\phi}{2}} |V \rangle_2 |V \rangle_3 \right),
\end{equation}
in which $\phi$ is the phase relation between the two contributions to the entangled state.

Photons 1 and 2 are sent to BS$_2$ and projected onto the maximally entangled state $|\Psi^- \rangle_{12}$ via a coincidence measurement between SPD$_1$ and SPD$_2$. Consequently, the reduced state reads
\begin{equation}
|\psi \rangle_{\rm coinc} \propto |\Psi^- \rangle_{12} \left( {\rm e}^{-{\rm i}\frac{\phi_1 - \phi}{2}} |H \rangle_3 - {\rm e}^{{\rm i}\frac{\phi_1 - \phi}{2}} |V \rangle_3 \right).
\end{equation}

For qubit 3, we first introduce a phase $\phi_3$ between the $|H\rangle_3$ and $|V\rangle_3$ components, and then rotate its polarization by $45^{\circ}$. In other words, we apply the following transformation:
\begin{eqnarray}
	|H \rangle_3 &\rightarrow& \frac{1}{\sqrt{2}} {\rm e}^{-{\rm i}\frac{\phi_3}{2}} \left(  |H \rangle_3 + |V \rangle_3 \right) \nonumber \\
	|V \rangle_3 &\rightarrow& \frac{1}{\sqrt{2}} {\rm e}^{{\rm i}\frac{\phi_3}{2}} \left(  |V \rangle_3 - |H \rangle_3 \right).
\end{eqnarray}

Projection is carried out by the f-PBS, followed by SPD$_3$ which measures the $|V\rangle_3$ component (see also \figurename~\ref{fig_Exp} for more details).
The threefold coincidence probability, $p_{123}$, between detectors SPD$_1$, SPD$_2$ and SPD$_3$, is then given by
\begin{equation}
 p_{123} \propto \sin^2 \left( \frac{\phi_1 - \phi + \phi_3}{2}\right).
\end{equation}
This probability function shows the required nonlocal relation between the phase settings at the qubit generator ($\phi_1$) and analysis stages ($\phi_3$). For our proof-of-principle experiment, we choose to keep $\phi_1$ and $\phi_3$ constant (at zero) and to vary $p_{123}$ by tuning the phase $\phi$ in the entangled state. The results in \figurename~\ref{fig_results} are in perfect agreement with the theoretical framework.

\section*{Acknowledgment}

The authors acknowledge the Agence Nationale de la Recherche for the e-QUANET and CONNEQT projects (grants ANR-09-BLAN-0333-01 and ANR-EMMA-002-01, respectively), the French Minist\`ere de l'Enseignement Sup\'erieur et de la Recherche (MESR), la Direction G\'en\'erale de l'Armement (DGA), the Malaysian government (MARA), the Conseil R\'{e}gional PACA, the iXCore Science Foundation, and the Foundation Simone \& Cino Del Duca (Institut de France), for financial support.
The authors also thank O. Alibart, A. Martin, L. Labont\'e, and V. D'Auria for fruitful discussions.

\end{document}